\documentclass[aoas,MSNbibl,nameyear,dvips]{arximspdf}
\usepackage{graphicx}
\doi{10.1214/12-AOAS598} 
\volume{7}
\issue{2}
\pubyear{2013}
\firstpage{669}
\lastpage{690}

\makeatletter

\def\cal{\mathcal}

\makeatother

\begin{document}
\begin{frontmatter}

\title{Refining genetically inferred relationships using treelet
covariance smoothing\thanksref{T1}}
\runtitle{Treelet covariance smoothing}
\thankstext{T1}{Supported by National
Institute of
Mental Health Grants MH057881 and MH097849, ONR Grant N0014-08-1-0673
and NSF Grant DMS0943577. Funding support for the CIDR Visceral
Adiposity Study (Study accession number: phs000169.v1.p1) was provided
through the Division of Aging Biology and the Division of Geriatrics
and Clinical Gerontology, NIA. The CIDR Visceral Adiposity Study
includes a genome-wide association study funded as part of the Division
of Aging Biology and the Division of Geriatrics and Clinical
Gerontology, NIA. Assistance with phenotype harmonization and genotype
cleaning, as well as with general study coordination, was provided by
Health ABC Study Investigators.}

\begin{aug}
\author[A]{\fnms{Andrew} \snm{Crossett}},
\author[B]{\fnms{Ann B.} \snm{Lee}\corref{}\ead[label=e1]{annlee@cmu.edu}},
\author[C]{\fnms{Lambertus} \snm{Klei}},
\author[C]{\fnms{Bernie}~\snm{Devlin}}
\and
\author[D]{\fnms{Kathryn} \snm{Roeder}}
\runauthor{A. Crossett et al.}
\affiliation{West Chester University, Carnegie Mellon University,
University of Pittsburgh School of Medicine,
University of Pittsburgh School of Medicine and Carnegie Mellon University}
\address[A]{A. Crossett\\
Department of Mathematics\\
West Chester University\\
West Chester, Pennsylvania 19383\\
USA} 
\address[B]{A. B. Lee\\
Department of Statistics\\
Carnegie Mellon University\\
5000 Forbes Ave, 229J Baker Hall\\
Pittsburgh, Pennsylvania 15213\\
USA\\
\printead{e1}}
\address[C]{L. Klei\\
B. Devlin\\
Department of Psychiatry\\
University of Pittsburgh School of Medicine\\
Pittsburgh, Pennsylvania 15213\\
USA}
\address[D]{K. Roeder\\
Department of Statistics\\
Carnegie Mellon University\\
Pittsburgh, Pennsylvania 15213\hspace*{7pt}\\
USA}
\end{aug}

\received{\smonth{3} \syear{2012}}
\revised{\smonth{8} \syear{2012}}

%
\begin{abstract}
Recent technological advances coupled with large sample sets have
uncovered many factors underlying the genetic basis of traits and the
predisposition to complex disease, but much is left to discover. A
common thread to most genetic investigations is familial relationships.
Close relatives can be identified from family records, and more distant
relatives can be inferred from large panels of genetic markers.
Unfortunately these empirical estimates can be noisy, especially
regarding distant relatives. We propose a new method for denoising
genetically---inferred relationship matrices by exploiting the
underlying structure due to hierarchical groupings of correlated
individuals. The approach, which we call Treelet Covariance Smoothing,
employs a multiscale decomposition of covariance matrices to improve
estimates of pairwise relationships. On both simulated and real data,
we show that smoothing leads to better estimates of the relatedness
amongst distantly related individuals. We illustrate our method with a
large genome-wide association study and estimate the ``heritability''
of body mass index quite accurately. Traditionally heritability,
defined as the fraction of the total trait variance attributable to
additive genetic effects, is estimated from samples of closely related
individuals using random effects models. We show that by using smoothed
relationship matrices we can estimate heritability using
population-based samples. Finally, while our methods have been
developed for refining genetic relationship matrices and improving
estimates of heritability, they have much broader potential application
in statistics. Most notably, for error-in-variables random effects
models and settings that require regularization of matrices with block
or hierarchical structure.

\end{abstract}

%
\begin{keyword}
\kwd{Covariance estimation}
\kwd{cryptic relatedness}
\kwd{genome-wide association}
\kwd{heritability}
\kwd{kinship}
\end{keyword}

\end{frontmatter}

\section*{Introduction}

In the past decade tremendous progress has been made toward
understanding the genetic basis of disease. This challenging endeavor
has given rise to numerous study designs with a vast arsenal of
statistical machinery. A~common theme, however, is the pivotal role
played by familial relationships. Traditionally relationships are
encoded in pedigrees of known relatives [Thompson (\citeyear{thompson74,thompson75}),
\citet{boehnke97},
\citet{epstein00},
\citet{mcpeek00}], but for more
distantly related individuals, pedigree information can sometimes be
erroneous or difficult to obtain. Relatedness can also be calculated
from large panels of genetic markers [\citet{milligan03},
\citet{albers08},
\citet{anderson2007maximum},
\citet{browning08},
\citet{browning10},
\citet{purcell07},
\citet{daylinkage},
\citet{yang10}].
While this approach has greatly expanded the scope of inference for
relationships, empirical estimates are noisy, especially regarding
distant relatives.

The search for a disease gene begins with finding unusual sharing of
genetic material among individuals who share a trait (phenotype).
Linkage analysis involves the study of joint inheritance of genetic
material and phenotypes within relatives [\citet{hopper82},
\citet{almasy98}].
Typically, these studies are restricted to relatives within a pedigree,
but more recently the approach has been extended to samples of people
who are more distantly related and without known pedigree structure
[\citet{daylinkage}]. Alternatively, genetic associations can be
discovered from population samples, which are usually based on
case--control studies. In these studies the sample is assumed to be
unrelated, but the presence of distant relatives (i.e., cryptic
relatedness) can reduce power or generate spurious associations [\citet{lander94},
\citet{astle2009population}]. Numerous methods have been proposed to
deal with familial structure in genetic association studies [\citet{choi09},
\citet{bravo09},
\citet{thornton10},
\citet{kang2010variance}], all of which require an
estimate of family relationships among individuals within the study.

Relationships are also critical for quantitative genetics. A common
problem for quantitative genetics is to estimate the fraction of
variance of a continuous trait, such as height, due to genetic
variation amongst individuals in a population. This feature, known as
heritability, delineates the relative contributions of genetic and
nongenetic factors to the total phenotypic variance in a population.
Heritability is a fundamental concept in genetic epidemiology and
disease mapping. Using a variety of close relatives, the heritability
of quantitative and qualitative traits can be estimated directly [\citet{fisher18},
\citet{devlin1997heritability}]. With complex pedigrees, applying
the same principles, heritability can be estimated using random effects
models [\citet{henderson1950estimation}]. Heritability of height,
weight, IQ and many other quantitative traits have been investigated
for nearly a century and continue to generate interest [\citet
{deary2012genetic}].

Interest in the genetic basis of disease is high because greater
understanding of disease etiology will in principle lead to better
treatments. Large population-based samples are enhancing our ability to
identify DNA variants affecting risk for disease and it has become the
standard to search for genetic variants associated with common disease
using genome-wide association studies (GWAS). Thousands of associations
for common diseases/phenotypes have already been validated [\citet
{visscher2012five}]. Nevertheless, even in the most successful cases,
such as Inflammatory Bowel Disease studied in \citet{mcgovern2010genome}
and \citet{imielinski2009common}, discoveries account for only a
fraction of the heritability.

Given the relatively limited discoveries thus far, a reasonable
question is whether the heritability of a trait estimated from
relatives truly does trace to genetic variation. \citet{yang10} offer a
novel approach to genetic analysis that shows that indeed much of it
does. They propose to analyze population samples, rather than
pedigrees, for the heritability of the trait. To do so they first
estimate the correlation between all pairs of individuals in the
population sample using a dense set of common genetic variants, such as
those typically used for a GWAS. They then take this matrix and relate
it to the covariance matrix of phenotypes for these subjects to derive
an estimate of heritability. Thus, in their application, where
essentially all relatives are removed from the sample, heritability
refers to the proportion of variance in the trait explained by the
measured genetic markers. They provide a fascinating example of how
this approach works in the case of human height and they and others
applied these techniques to many other traits [reviewed by \citet
{visscher2012five}].

The work of Yang et al. (\citeyear{yang10})
inspired us to consider applying a related
approach to answer a different question. Could estimates of relatedness
obtained from a population sample be improved by using smoothing
techniques on the variance--covariance matrix? If so, population samples
could be used to estimate heritability---in the traditional sense---without requiring close relatives. This approach has application to
phenotypes for which extended pedigrees are difficult to obtain. For
instance, there is controversy in the literature concerning the
heritability of autism, which is typically estimated from twin studies
[\citet{hallmayer2011autism}]. Smoothing techniques could also be used
to estimate relatedness in samples of distantly related individuals for
many other genetic analyses. For example, a version of linkage analysis
could be applied to distant relatives.

We propose Treelet Covariance Smoothing---a novel method for smoothing
and multiscale decomposition of covariance matrices---as a means to
improving estimates of relationships. Treelets were first introduced in
\citet{lee2007treelets} and \citet{lee08} as a multi-scale basis that
extends wavelets to unordered data. The method is fully adaptive. It
returns orthonormal basis functions supported on nested clusters in a
hierarchical tree. Unlike other hierarchical methods, the basis and the
tree structure are computed simultaneously, and both reflect the
internal structure of the data.

In this work, we extend the original treelet framework for smoothing of
one-dimensional signals to smoothing and denoising of
variance--covariance matrices with hierarchical block structure and
unstructured noise. Smoothing is achieved by a nonlinear approximation
scheme in which one discards small elements in a multi-scale matrix
decomposition. The basic idea is that if the data have underlying
structure in the form of groupings of correlated variables, then we can
enforce sparsity by first transforming the data into a treelet
representation by a series of rotations of pairs of correlated
variables, and then thresholding covariances. We refer to this new
regularization approach for covariance matrices with groupings on
multiple scales as \textit{Treelet Covariance Smoothing} (TCS).

We apply TCS to genetically inferred relationship matrices, with the
goal of improving estimates of pairwise relationships from large
pedigrees and population-based samples. On both simulated and real
data, we show that TCS leads to better estimates of the relatedness
between individuals. Using these estimates allows us to estimate the
heritability from population-based samples provided they include some
distantly related individuals, a property that is almost inevitable in practice.
Finally, we discuss how estimating heritability is simply a case of
variance component estimation for an error-in-variables random effects
model. Therefore, our method can be applied to a whole family of more
general models of similar structure.

\section*{Models}

\subsection*{GWAS panels}
The human genome contains many millions of single nucleotide
polymorphisms (SNPs) and other genetic variation distributed across the
genome. In a GWAS it is now typical to measure a panel of at least
500,000 SNPs from each subject. SNPs typically have only two forms or
alleles within a population. Whichever allele is less frequent is
called the minor allele. The genotype of an individual at a SNP can
then be coded as 0, 1 or 2 depending on the number of minor alleles the
individual has at that SNP. Alleles at SNPs in close physical proximity
are often highly correlated (i.e., in linkage disequilibrium). When
multiple SNPs are in linkage disequilibrium, we say one of these SNPs
``tags,'' or represents, the others. Although estimates vary,
well-designed panels of 500,000 SNPs do not tag all of the common SNPs
in the genome and they tag very few of the SNPs with rare minor alleles
[\citet{yang10}]. Nevertheless, GWAS provide considerable information
about familial relationships.

\subsection*{Estimating genetic relationships}
The relatedness between a pair of individuals is defined by the
frequency by which they share alleles \textit{identical by descent} (IBD).
Formally, two alleles are considered IBD if they descended from a
common ancestor without an intermediate mutation. Within a pedigree
relatives share very recent common ancestors, hence, many alleles are
IBD. For a more detailed exposition of genetic relationships, see \citet
{astle2009population}.

The quantity of interest in this investigation is the \textit{Additive
Genetic Relationship} which is defined as the expected proportion of
alleles IBD for a pair of individuals. For individuals $i$ and $j$ we
use $A_{ij}$ to denote this quantity, which is more familiar when
viewed as the \textit{degree of relationship}, where $R_{ij}=-\log
_2(A_{ij})$. For example, for siblings, first cousins and second
cousins, who are 1st, 3rd and 5th degree relatives, $A$ is $1/2$, $1/8$ and
$1/32$, respectively. Within a noninbred pedigree $A$ can be computed
using a recursive algorithm [\citet{thompson1986pedigree}]. For example,
if individual $i$ has parents $k$ and $l$, then $A_{ij}=A_{ji} =
1/2(A_{jk}+A_{jl})$.

For distantly related individuals, detailed pedigree information is not
often available; however, with GWAS data one can calculate
genome-average relatedness directly [\citet{astle2009population}]. Even
with complete information regarding IBD status of the chromosomes, the
fraction of genetic material shared by relatives will differ slightly
from the expectation calculated from the pedigree due to the stochastic
nature of the meiotic process [\citet{weir2006genetic}]. For the purpose
of genetic investigations, one could argue that genome-average
relatedness is a truer measure of relatedness. For example, while two
distantly related individuals are expected to share a small fraction of
their genetic material, if they do not inherit anything from their
common ancestor, it seems appropriate to consider them unrelated.

Under many population genetic models $A_{ij}$ can also be interpreted
as a correlation coefficient. Let $z_{ik}$ denote the scaled minor
allele count for individual $i$ at SNP $k$: $z_{ik} =
(z_{ik}^*-2p_k)/(2p_k(1-p_k))^{1/2}$, where $z_{ik}^*$ is the minor
allele count and $p_k$ is the minor allele frequency. For individuals
$i$ and $j$ at genetic variant $k$, it follows from our model that
%
\begin{equation}
\operatorname{Cov}[z_{ik},z_{jk}] = A_{ij}.
\label{eqSNPcov}
\end{equation}

Exploiting this feature leads to a method of moments estimate of $A$
from a panel of $m$ genetic markers. To see this, let $\mathbf{z}_k$
denote a column vector of observed scaled allele counts for all
individuals at the $k$th SNP, then let
%
\begin{equation}
\hat A = \frac1m\sum_{k=1}^m
\mathbf{z}_k \mathbf{z}_k^t =
\frac{Z Z^t} m, \label{eqZZprime}
\end{equation}
where $Z= (\mathbf{z}_1, \ldots, \mathbf{z}_m)$. The Genome-wide
Complex Trait Analysis (GCTA) software from \citet{yang2010gcta}
computes this estimator.

The method of moments estimator is unbiased if the population allele
frequencies are known [\citet{milligan03}]. In practice, the $p_k$'s are
estimated from the sample data. A criticism of this estimator is that
some off-diagonal elements are negative, which does not conform to the
interpretation of $A_{ij}$ as a probability. Viewed as a correlation
coefficient, however, negative quantities suggest the pair of
individuals share fewer alleles than expected given the allele
frequencies. Alternatively,
maximum likelihood estimators of $A$ have been developed [\citet
{thompson75},
\citet{milligan03}], but these estimators are quite
computationally intensive for GWAS panels. Hence, while method of
moments estimators are typically less precise than maximum likelihood
estimators, they are more commonly used when a large SNP panel is available.

\subsection*{Estimating heritability}

By definition, the heritability of a quantitative trait ($y$) such as
height is determined by the additive effect of many genes and genetic
variants ($g$), each of small effect (i.e., the polygenic model). For
individuals $i=1,\ldots,n$, suppose that the genetic effects are
explained by $J$ causal SNPs, and we can express the genetic effect as
%
\begin{equation}
g_i = \sum_{j=1}^Jz_{ij}u_j,
\label{eqSNPs}
\end{equation}
where $u_j$ is the additive random effect of the $j$th causal variant,
weighted by the scaled number $z_{ij}$ of minor alleles at this
variant. Let $\mathbf{g}=(g_1,\ldots,g_n)^t$ be the vector of random
effects corresponding to the additive genetic effects for individuals
$i=1,\ldots,n$. For $\mathbf{u}=(u_1,\ldots,u_J)^t$ and $Z_c=
[z_{ij}]$, we write $\mathbf{g}= Z_c\mathbf{u}$. Define $G$ as the
variance--covariance matrix of $\mathbf{g}$. Assuming $\operatorname{Var}[\mathbf
{u}]= I\sigma^2_u$, it follows that
%
\begin{equation}
G = {\sigma^2_g} \frac{Z_cZ_c^t}{J}, \label{eqdual}
\end{equation}
where $\sigma^2_g = J\sigma^2_u$.

In the traditional model for quantitative traits a continuous phenotype
$y$ is modeled as
%
\begin{equation}
y_i = \mu+ g_i + e_i, \label{eqqtmodel}
\end{equation}
where $\mathbf{e}=(e_1,\ldots,e_n)^t$ is the vector of residual
effects, and $\mathbf{y}=(y_1,\ldots,y_n)^t$ is the vector of
phenotypes. In matrix notation, $\mathbf{y} = \mathbf{1}\mu+ \mathbf
{g} + \mathbf{e}$. The residuals are assumed to be independent with
variance--covariance equal to $I \sigma^2_e$ and the random effects and
residual error are assumed to be normally distributed.
Consequently,
%
\begin{equation}
\operatorname{Var}[\mathbf{y}] = \frac{Z_cZ_c^t}{J}\sigma^2_g + I
\sigma_e^2. \label{eqknownZc}
\end{equation}
The heritability of the phenotype $y$ is defined as
\[
h^2 = \frac{\sigma_g^2}{\sigma_g^2+\sigma_e^2}.
\]
This quantity is more accurately known as the additive or narrow-sense
heritability, in contrast to the broad-sense heritability, which
includes nonadditive genetic effects such as gene--gene interactions.
Our inferences will be confined to narrow-sense heritability.

If the causal SNPs (or good tag SNPs) and the phenotype were directly
measured, then one could estimate $h^2$ based on equation (\ref
{eqqtmodel}) and the implied random effects model using maximum
likelihood (REML)
[\citet{searle1992variance}]. Notationally, $Z_c$ is an $n\times J$
matrix that picks out $J$ columns of the full SNP panel $Z$. In
practice, $Z_c$ is not known. Few of the causal SNPs are known for any
phenotype, and many causal SNPs will be missing from $Z$ (i.e., not
tagged by any measured SNPs).

How then is $h^2$ estimated in practice? Assuming various subsets of
individuals in the sample are related with relationship matrix $A$
(defined previously), heritability can be estimated without any
knowledge of causal genetic variants that constitute $\mathbf{g}$. From
equation~(\ref{eqSNPcov}) and the polygenic model it follows that $\frac
{Z_cZ_c^t}{J}\to A$ as $J$ gets large. This inspires an alternative
random effects model which has long been utilized in population genetics:
%
\begin{equation}
\operatorname{Var}[\mathbf{y}] = A\sigma^2_g + I
\sigma_e^2. \label{equnknownZc}
\end{equation}
Historically, $A$ has been derived from known pedigree structure.
However, provided some subsets of the individuals in the sample are
related (even distantly), one can estimate $A$ from genetic markers
using either method of moments or maximum likelihood estimation
techniques. This approach has been applied frequently in quantitative
genetics, especially in breeding studies [\citet{lynch1999estimation},
\citet{eding2001marker},
\citet{visscher2006assumption},
\citet{hayes2008technical}].
We conjecture that by using TCS, we can improve estimates of $A$ and
obtain better estimates of heritability without knowledge of causal variants.

Alternatively, if the sample is completely unrelated, then substituting
the result of equation~(\ref{eqZZprime}) for (\ref{eqknownZc}) does not
lead to an estimate of $h^2$ unless all of the causal SNPs have been
recorded. Instead this approach estimates $h^2_{s}\le h^2$, the
proportion of the variance in phenotype explained by the SNP panel
[\citet{yang10}]. In this setting, TCS will not improve estimates of $h^2_s$.

\section*{Methods}

\subsection*{Treelet covariance smoothing (TCS)} \label{secTCS}

The genetic relationship matrix $A$ is a measure of the additive
covariance structure that exists between individuals due to a common
genetic background. We estimate the relationship matrix using genotyped
SNPs, but this estimate is usually noisy. Hence, we propose a method
for improving upon this estimate using treelets.

Treelets simultaneously return a hierarchical tree and orthonormal
basis functions supported on \textit{nested clusters} in the tree---both
reflect the underlying structure of the data. Here we extend the
original treelet framework [\citet{lee2007treelets},
\citet{lee08}] for smoothing
one-dimensional signals and functions, to a new means of smoothing and
denoising variance--covariance matrices with hierarchical block
structure and unstructured noise. The main idea is to first move to a
different basis representation through a series of local
transformations, and then impose sparsity by thresholding the
transformed covariance matrix. We refer to the approach as Treelet
Covariance Smoothing (TCS). The general setup is as follows. [See
Appendix in \citet{lee08} for details on how to compute the treelet
transformation. The treelet algorithm, as well as its implementation,
is available in R on CRAN as the \texttt{treelet} library.]

Let $\mathbf{z}$ be a random vector in $\mathbb{R}^N$ with
variance--covariance matrix $\Sigma$. In our context, $\mathbf{z}$
represents the scaled minor allele counts for a set of $N$ individuals
at any SNP, and the covariance $\Sigma=A$, the additive genetic
relationship matrix of the $N$ individuals [equation~(\ref{eqSNPcov})].
Now at each level of the treelet algorithm, we have an orthonormal
multiscale basis. Let $\{\mathbf{v}_1,\ldots,\mathbf{v}_N\}$ denote the
basis at the top of the tree [corresponding to level $\ell= N-1$ if
using the notation in \citet{lee08}]. We write\vspace*{-2pt}
%
\begin{equation}
\mathbf{z} = \sum_{i=1}^Nc_i
\mathbf{v}_i, \label {eqtreeletdecomp}\vspace*{-1pt}
\end{equation}
where $c_i = \langle\mathbf{z},\mathbf{v}_i \rangle$
represent the orthogonal projections onto local basis vectors on
different scales. It follows that the covariance of $\mathbf{z}$ can be
written in terms of a \textit{multi-scale matrix decomposition}\vspace*{-1pt}
%
\begin{equation}
\Sigma= \operatorname{Var}(\mathbf{z}) = \sum_{i=1}^N
\gamma_{i,i} \mathbf {v}_i(\mathbf{v}_i)^t
+ \sum_{i\ne j}^N\gamma_{i,j}
\mathbf{v}_i(\mathbf {v}_j)^t, \label{eqcov}\vspace*{-1pt}
\end{equation}
where $\gamma_{i,i} = \operatorname{Var}(c_i)$ and $\gamma_{i,j} =
\operatorname{Cov}(c_i,c_j)$. The first term in equation~(\ref{eqcov})
describes the diagonally symmetric block structure of the
variance--covariance matrix. These blocks are organized in a
hierarchical tree. The second term describes a more complex structure,
including off-diagonal rectangular blocks, which are also
hierarchically related to each other in a multi-scale matrix decomposition.

In practice, the covariance $\Sigma$ is unknown, and both the
covariance matrix and the treelet basis need to be estimated from data.
For relationship matrices, one can, for example, derive an estimate
$\widehat{\Sigma} = \widehat{A}$ from marker data using method of
moments or maximum likelihood methods.
Denote the treelet basis derived from $\hat\Sigma$ by $\{\widehat
{\mathbf{v}}_1,\ldots, \widehat{\mathbf{v}}_N\}$, and write\vspace*{-1pt}
\[
\widehat{\Sigma} = \sum_{i=1}^N\widehat{
\gamma}_{i,i}\widehat{\mathbf {v}}_i(\widehat{
\mathbf{v}}_i)^t + \sum_{i\ne j}^N
\widehat{\gamma }_{i,j}\widehat{\mathbf{v}}_i (\hat{
\mathbf{v}}_j)^t,\vspace*{-1pt}
\]
where $\widehat{\gamma}_{i,i} = \widehat{\operatorname{Var}}(c_i)$
and $\widehat{\gamma}_{i,j} = \widehat{\operatorname{Cov}}(c_i,c_j)$.

Let $T(\widehat{\Sigma})$ be the covariance estimate after a treelet
transformation, that is, after applying a full set of $N-1$ Jacobi
rotations of pairs of correlated variables. A~calculation shows that\vspace*{-1pt}
%
\begin{equation}
\widehat{\gamma}_{i,i} = \widehat{\operatorname{Var}}(c_i) =
\bigl[T(\widehat {\Sigma})\bigr]_{ii}  \qquad  \mbox{and}\qquad    \widehat{
\gamma}_{i,j} = \widehat {\operatorname{Cov}}(c_i,c_j)
= \bigl[T(\widehat{\Sigma})\bigr]_{ij}, \label{eqtreeletcoeffs}\vspace*{-1pt}
\end{equation}
where $c_i \equiv\langle\mathbf{z},\mathbf{v}_i \rangle$ and $c_j
\equiv\langle\mathbf{z},\mathbf{v}_j \rangle$. This suggests\setcounter{footnote}{1}\footnote
{The special case $c_i \equiv\langle\mathbf{z},\delta_i \rangle$ and
$c_j \equiv\langle\mathbf{z},\delta_j \rangle$, where $\delta_i$
denotes the Kronecker delta function, corresponds to simple
thresholding of the original covariance estimate. Here we consider more
general groupings of correlated variables on different scales.} a
smoothed estimate of the covariance by thresholding:\vspace*{-1pt}
%
\begin{equation}
\widetilde{\Sigma}(\lambda) = \sum_{i=1}^Nf_\lambda[
\widehat{\gamma }_{i,i}]\widehat{\mathbf{v}}_i(\widehat{
\mathbf{v}}_i)^t + \sum_{i\ne
j}^Nf_\lambda[
\widehat{\gamma}_{i,j}]\widehat{\mathbf{v}}_i(\widehat {
\mathbf{v}}_j)^t, \label{eqcovsmooth}
\end{equation}
with the thresholding function
%
\begin{equation}
f_\lambda[a] = \cases{ %
a, &\quad $\mbox{when $|a|
\ge\lambda$},$
\vspace*{2pt}\cr
0, &\quad $\mbox{when $|a|< \lambda$,}$}
\label{eqcovthresholding}
\end{equation}
where $\lambda$ is a smoothing parameter.

To summarize and in matrix short-hand notation, the smoothed genetic
relationship matrix is given by
%
\begin{equation}
\widetilde{A}(\lambda) = B f_{\lambda}\bigl[T(\widehat{A})\bigr]
B^t, \label{eqAsmoothmat}
\end{equation}
where $B= (\hat{\mathbf{v}}_1,\ldots,\hat{\mathbf{v}}_N)$ and $T(\widehat
{A})$, respectively, denote the treelet basis and the covariance matrix
at the top of the tree, and $f_\lambda$ corresponds to element-wise
thresholding [equation~(\ref{eqcovthresholding})]. Note that to compute
$B$ we only need to know the Jacobi rotations at each level of the
tree, more precisely, the treelet basis, $B = J^{(1)} \cdot J^{(2)}
\cdot\cdots\cdot J^{(N-1)}$, where the Jacobi rotation matrix
$J^{(\ell)}$ is the rotation matrix at level $\ell$. The covariance
estimate after a treelet transformation and before smoothing is $\widehat
{\Sigma}^\ell\equiv T(\widehat{A}) = B^t \widehat{A} B$.

\subsection*{Choosing a smoothing parameter}

The goal is to choose a threshold ($\lambda$) that reduces noise in the
estimated relationships. Traditional cross-validation is not an option
because we cannot predict $A_{ij}$ without including persons $i$ and
$j$. Alternatively, we have an abundance of genetic information from
which to estimate $\hat{A}$. We propose a SNP subsampling procedure to
estimate the tuning parameter.

We begin by breaking the genome into independent \textit{training} and
\textit{test} sets by randomly placing half the chromosomes into each set.
To improve the efficiency of our estimate of $A$, we utilize a
``blackout window'' of length $b$ to avoid sampling SNPs that are
highly correlated. This $b$ can be considered either in terms of
physical location along the chromosome or the number of SNPs between
any two SNPs in question. From the set of training chromosomes, select
a relatively large sample of $M$ independent SNPs to get a reliable
estimate of $\hat{A}$. We train our algorithm
by smoothing $\hat{A}$ using TCS to get $\widetilde{A}(\lambda)$, for
all $\lambda\in\Lambda$, where $\Lambda$ is a grid of reasonable
threshold values.

Once we have $\widetilde{A}(\lambda)$, for a given $\lambda$, we
subsample $L$ SNP sets of size $k$ from the test set of chromosomes.
Here, $k\ll M$ and the SNPs within each of the $L$ subsampled sets
follow our defined blackout window, $b$. Then, for all $l = 1,\ldots,L$, estimate the relationship matrix, $\hat{A}_l$, based on the subset
of SNPs. We then compare our smoothed relationship matrix, $\widetilde
{A}(\lambda)$, from the training chromosomes to each of the $L$
nonsmoothed relationship matrices, $\hat{A}_l$, via a weighted risk function:
%
\begin{equation}
\label{eqpicklambda} H(\lambda) = \frac{1}{(N-1)NL} \sum
_{l=1}^{L}\sum_{i<j}^{N}
w_{ij} \bigl(\hat{A}_{ij,l} - \widetilde{A}_{ij}(
\lambda) \bigr)^2,
\end{equation}
where $w_{ij}$ is a weight associated with each element in
$A$. Clearly, the optimal tuning parameter is $\hat{\lambda} = \operatorname
{argmin}_{\lambda\in\Lambda}H(\lambda)$.

The reason for introducing the weighting scheme is because many
subjects are nearly unrelated. Thus, we aim to upweight the loss
function so that the preponderance of near-zero elements in the
off-diagonal do not overwhelm the loss function. We suggest using the
learned hierarchical tree to get the weights. More specifically,
$w_{ij} = |[T(\hat{A})]_{ij}|$, corresponding to the absolute value of
the correlations between the final groupings of individuals after $N-1$
rotations [equation~(\ref{eqtreeletcoeffs})]. Also, we set $w_{ii} = 0$
because we are not interested in estimating inbreeding coefficients. It
should be noted that this is a rather general weighting method. Other
schemes may be more appropriate if there is a priori information
suggesting the importance of particular relationships.


\section*{Results}
\subsection*{Simulations}

\begin{figure}

\includegraphics{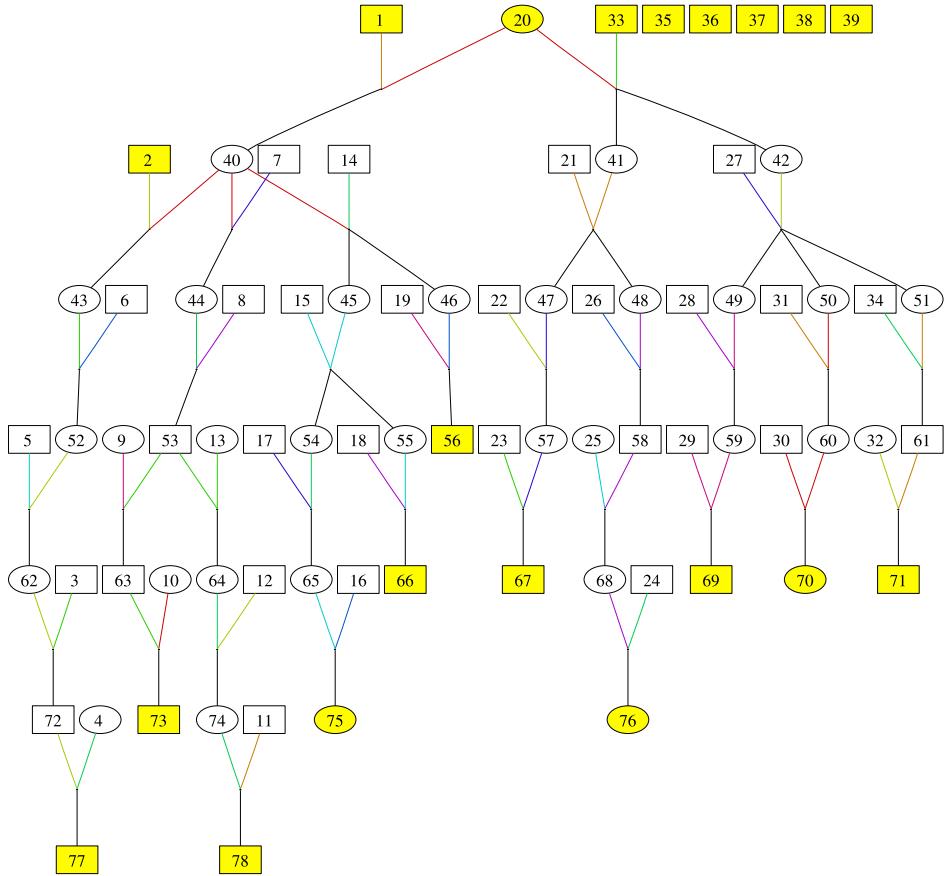}

\caption{Pedigree of a single family used for simulations. Genomes were
dropped through the entire pedigree and ten individuals were sampled
from the twenty possible highlighted individuals. Individuals 35--39 are
unrelated to everyone else in the pedigree.}
\label{figpedigree}
\end{figure}

To produce realistic simulations, we started with the phased genomes
(haplotypes) of individuals from the HapMap 3 database\footnote{\url
{http://hapmap.ncbi.nlm.nih.gov/downloads/phasing/2009-02_phaseIII/HapMap3_r2/}},
selecting two populations with European ancestry (CEU and TSI).
Utilizing the small sample of available haplotypes, our first objective
was to generate a large sample of haplotypes, representative of those
that might be sampled from unrelated founders of a population. The
challenge was to keep intact the realistic haplotype structure for
a human population, including linkage disequilibrium (LD), without
generating unusual sharing between the founders. To accomplish this
goal, we took the HapMap data on CEU and Tuscan samples, which were
phased quite accurately into haplotypes, as the initial sample of
chromosomes from which to generate founders. Now each founder
haplotype was created by sampling pieces of chromosomes (or
haplotypes) from the initial sample. To do so, the number of
recombination spots per chromosome was determined using an overall
recombination of $\theta= 10^{-6}$ per Mb, which is 100 times the
normal rate of recombination for humans. The actual location of the
recombination spots were then determined using the recombination map
provided by HapMap, a procedure that successfully keeps intact the LD
structure of the chromosome. From this pool of generated haplotype
pairs, chromosomes were randomly assigned to each of the 39 founders
in each of 100 families. These founder chromosomes were then dropped
through a seven generation pedigree; see Figure \ref{figpedigree} for the pedigree of
a single family used for simulations. At each generation the
chromosomes underwent recombination with an overall rate of $\theta=
10^{-8}$ at locations determined by HapMap's recombination map. Within
each pedigree, the genotype information of twenty individuals was
collected (colored yellow). We then sampled ten individuals of varying
relatedness from this group with a random sampling strategy that
favored individuals of distant relatedness within the pedigree. The
highest pairwise relatedness within a family is 0.125, corresponding to
$R = 3$, and the lowest is $<0.001$. Individuals from different families
are unrelated. Each simulation produced a total of 1000 individuals
made up of 100 ten-member families of varying levels of relatedness.
Finally, the entire process was repeated fifty times.

\begin{figure}

\includegraphics{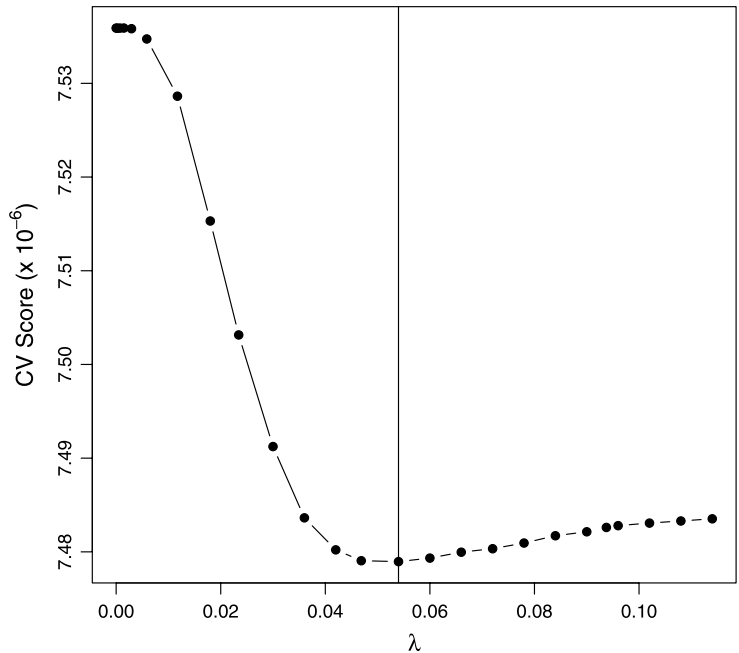}

\caption{Cross-validation plot showing the weighted risk function at
varying levels of the thresholding parameter, $\lambda$. The optimal
$\lambda$ is the point where the $H(\lambda)$ (CV Score) is minimized.}
\label{figcv}
\end{figure}

Because we know the pedigree structure, we can compare the unsmoothed
estimate $\hat{A}$ to $\tilde{A}$ found via TCS. Here, we use the GCTA
software [\citet{yang2010gcta}] to estimate $\hat{A}$ using 100,000
randomly chosen SNPs with MAF $> 0.05$. The optimal level of smoothing
($\hat{\lambda}$) is chosen via the subsampling scheme described
previously using $M=5000$, $b = 10$, $k = 50$, $L = 50$ and repeating
everything ten times. Here, $b$ is in terms of number of SNPs. We
choose $\hat{\lambda}$ by examining a plot of $H(\lambda)$ across a
grid of $\lambda$ values. The optimal smoothing parameter is the one
that minimizes the risk function, $H$. For one such simulation sample
we can see from Figure~\ref{figcv} that $\hat{\lambda}\approx0.051$.

\begin{figure}

\includegraphics{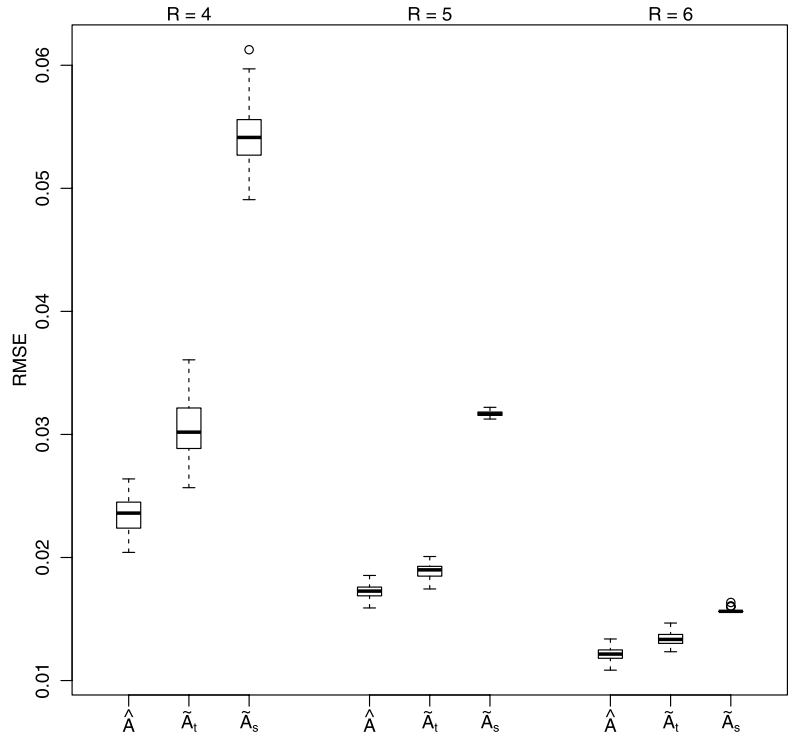}

\caption{Boxplots of RMSE for unsmoothed ($\hat{A}$) along with
smoothed using TCS ($\tilde{A}_{\mathrm{t}}$) and simple thresholding
($\tilde{A}_{\mathrm{s}}$) at increasing degrees of relatedness
($R=4,5,6$; see header). Here, TCS is better than simple thresholding
as the latter method thresholds too aggressively.}
\label{figrmse46}
\end{figure}

\begin{figure}

\includegraphics{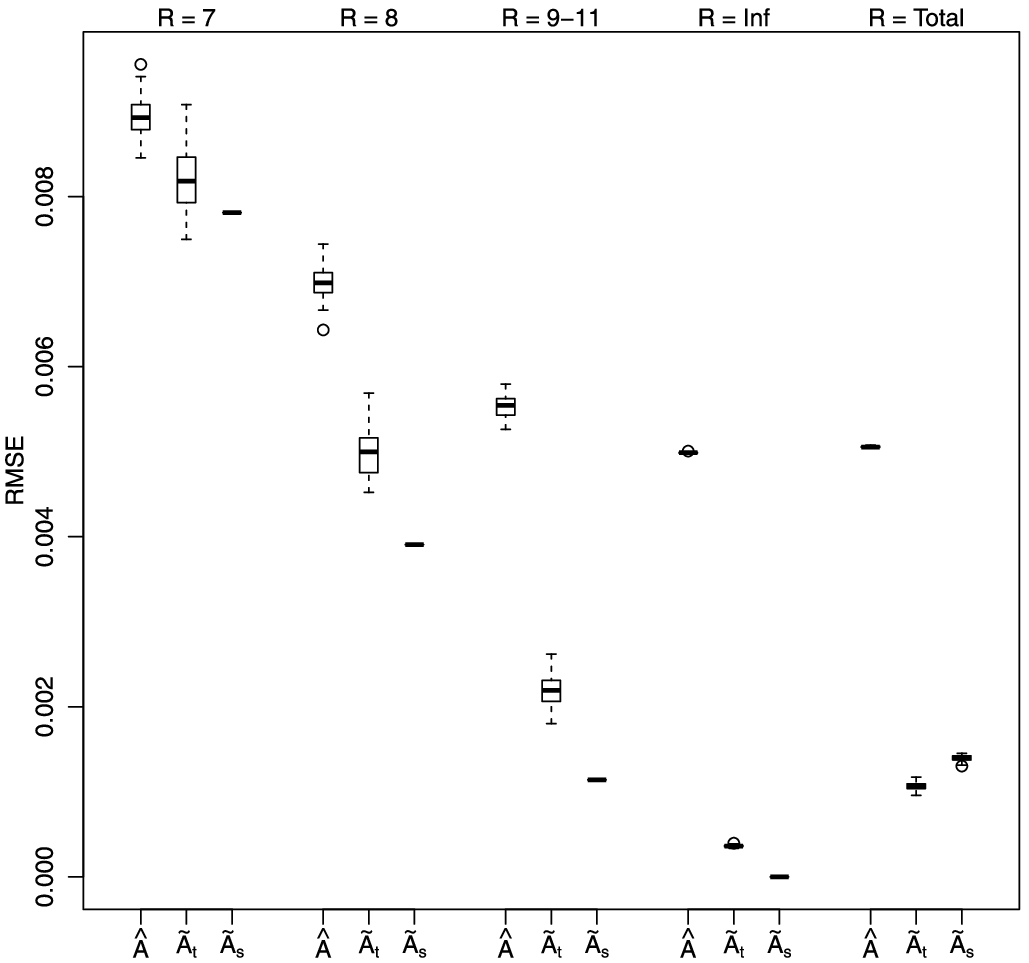}

\caption{Boxplots of RMSE for unsmoothed ($\hat{A}$) along with
smoothed using TCS ($\tilde{A}_{\mathrm{t}}$) and simple thresholding
($\tilde{A}_{\mathrm{s}}$) at increasing degrees of relatedness
($R=7,8,9\mbox{--}11$). Also included is the comparison of RMSE values for
unrelated pairs ($R=\operatorname{Inf}$) and average RMSE for the entire
relationship matrix ($R = \operatorname{Total}$). We see that both
thresholding methods remove noise, but TCS works better than simple
thresholding overall.}
\label{figrmse7T}
\end{figure}

\begin{figure}

\includegraphics{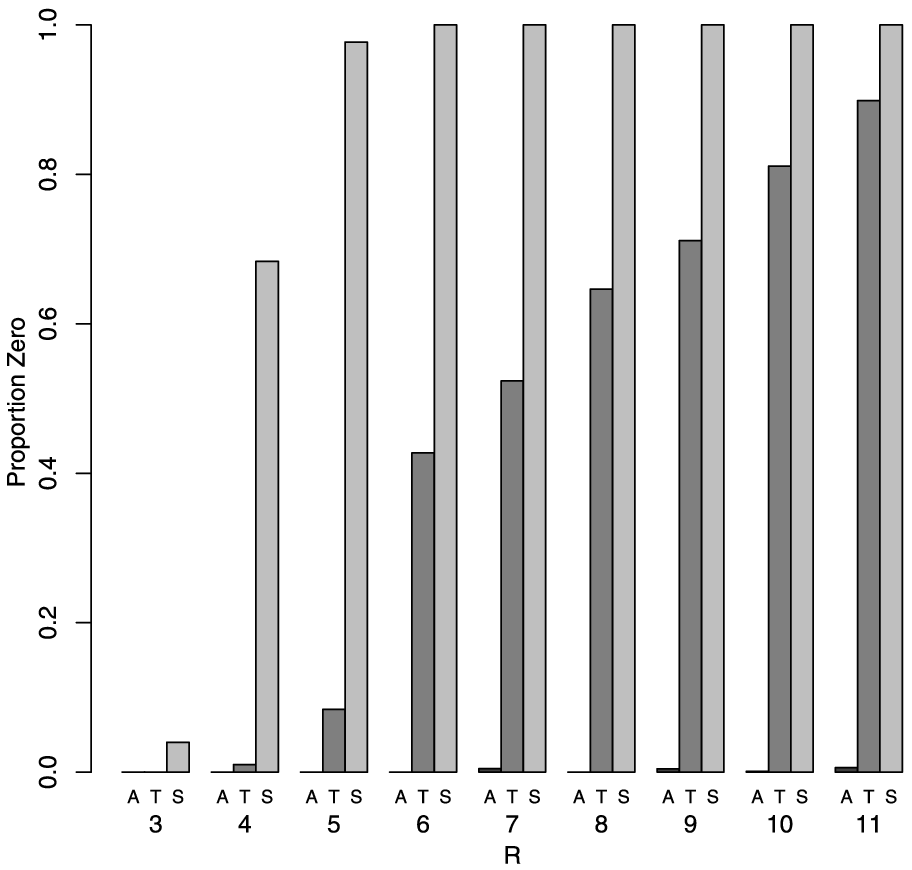}

\caption{Barplots of the percentage of relationships that are equal to
0 for no smoothing ($A$), smoothing using TCS ($T$) and simple
thresholding ($S$). The three cases are compared at increasing degrees
of relatedness ($R=3,\ldots,11$). Any value below $\varepsilon= 10^{-5}$
is considered to be 0.}
\label{figprop0}
\end{figure}

The question then becomes, does TCS improve estimates of relatedness?
Figures~\ref{figrmse46} and \ref{figrmse7T} display boxplots
comparing the root mean square error (RMSE) of $\hat A$ to $\tilde{A}$
at varying levels of known pairwise relationship values. For a full
comparison, we have included two smoothing methods: TCS, as previously
described, as well as ``simple thresholding,'' wherein the elements of
$\widehat A$ are directly thresholded. [The latter approach is a
degenerate case of TCS models, at level $\ell=0$, for which the basis
is the Dirac basis, i.e., $\mathbf{v}_i=\hat{\mathbf{ v}}_i = \delta_i$
for $i=1,\ldots, N$ in equations~(\ref{eqtreeletdecomp})--(\ref
{eqAsmoothmat}).] Moving from left to right in the figures, the true
degree of relatedness increases from $R=4,\ldots,11$, to no
relatedness. Over the entire matrix of estimates, the RMSE is 0.0055,
0.0015 and 0.0019 for the unsmoothed ($\hat{A}$), TCS ($\tilde{A}_{\mathrm
{t}}$) and simple thresholding ($\tilde{A}_{\mathrm{s}}$) methods
respectively, demonstrating an overall advantage of TCS. As with many
shrinkage methods, TCS introduces a slight bias that is reflected in a
higher RMSE for closely related individuals. Consequently, TCS has a
larger RMSE than the unsmoothed estimate for smaller values of $R$.
Where TCS gains a notable advantage over the unsmoothed estimate is in
differentiating between more distantly related individuals and noise.
From Figures~\ref{figrmse46} and~\ref{figprop0} we can see that
simple thresholding incurs a substantially larger RMSE for closer
relationships because it thresholds too aggressively. For $R=4$, 70\%
of the pairs are zeroed out, and for $R >4$ virtually all pairs are
zeroed out. Naturally, this method has the smallest RMSE for the sample
of unrelated pairs because thresholding zeros out all of these entries.
Notably, TCS does almost as well in this setting. A~direct comparison
of RMSE does not fully reflect the true loss incurred in practice. In
most genetic studies close relatives are often recorded in pedigrees
and, hence, estimates are not required. Alternatively, considering
distant relatives to be unrelated leads to a substantial loss for
estimating heritability and most other genetic applications.

\subsection*{Heritability in health ABC study}
Body Mass Index (BMI) is one of several traits measured as part of the
study entitled ``Whole Genome Association Study of Visceral Adiposity''
as part of the Health Aging and Body Composition (Health ABC) Study.
These data are archived in the Database for Genotypes and Phenotypes
(dbGaP)\footnote{\url{http://www.ncbi.nlm.nih.gov/gap}}. We restrict our
attention to those 1644 individuals with self-reported European
ancestry. To control for confounding, prior to analysis, we adjust BMI
scores by regressing out age, gender and collection site. Our objective
is to estimate heritability of BMI from this population sample.
Published heritability estimates range from as low as 0.05 to as high
as 0.90 [\citet{allison1996heritability}]; however, based on estimates
derived from known pedigrees, the heritability of BMI is estimated to
be approximately 50--75\% [\citet{kangas10},
\citet{zabaneh09}].

From the full sample of SNPs (Illumina 1M platform) we remove those
with missingness greater than 0.1\% and MAF $< 0.01$. From these we
select a subpanel of $90\mbox{,}000$ SNPs, chosen to be nearly evenly spaced.
Based on these SNPs, we calculated the relationship matrix $\hat A$,
and find that the individuals are predominately unrelated. The most
highly related pair appear to be third degree relatives, such as first
cousins. And more than half of the pairs appear to be more distantly
related than 10th degree relatives.

To estimate the heritability in this setting, we input the smoothed
relationship matrix in equation~(\ref{eqcovsmooth}) into the REML
algorithm. The required smoothing parameter $\lambda$ is selected in
two ways: (i) minimizing the loss function in equation~(\ref
{eqpicklambda}) via the subsampling approach; and (ii) using a
profile likelihood approach. With both techniques, we get estimates of
the heritability that are very close to what is found in the literature.

\begin{figure}

\includegraphics{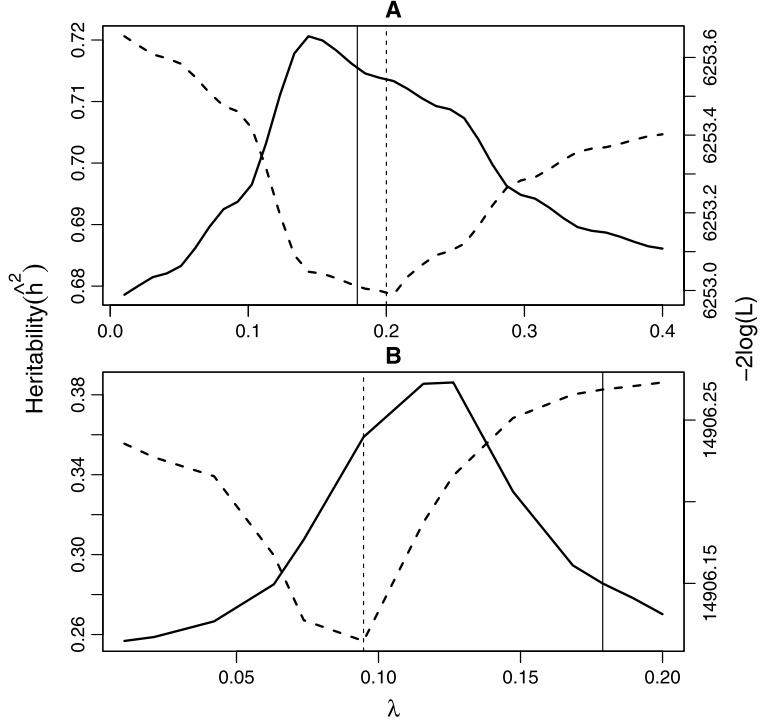}

\caption{Estimating heritability in the Health ABC data set. Solid
curve is the estimated heritability at increasing values of the
smoothing parameter $\lambda$. The dashed curve is $\propto-2\operatorname
{log}({\cal L})$, where, ${\cal L}$ is the maximum profile likelihood
obtained from the REML algorithm. The solid vertical line is the
optimally chosen threshold value using our subsampling scheme. The
dashed vertical line represents the optimally chosen threshold value
when minimizing the likelihood profile. \textup{A}: For BMI, $h^2 = 0.72$
when using subsampling to choose an optimal smoothing parameter $(\hat
{\lambda}=0.18)$. Similarly, $h^2 = 0.71$ when using the profile
likelihood plot $(\hat{\lambda}=0.20)$. With no smoothing $(\lambda=0)$,
$h^2 = 0.23$. This is not shown on the plot. \textup{B}: For AVFD, $h^2 =
0.29$ when using our subsampling approach to choose an optimal smoothing
parameter. However, $h^2 = 0.36$ when using the profile likelihood plot
$(\hat{\lambda}=0.09)$. These are compared to $h^2=0.11$, the
heritability when there is no smoothing (not shown).}
\label{figBMIAVFD}
\end{figure}

For a range of smoothing parameters, $0 \leq\lambda\leq0.40$, we
calculate the smoothed relationship matrix, $\tilde{A}_\lambda$, and
plug this value into the REML model to obtain a profile likelihood
(Figure~\ref{figBMIAVFD}). Also plotted in this figure is $\widehat
{h}^2_{\lambda}$, the heritability that maximizes REML as a function of
$\lambda$ (or minimizes---2 times the log-likelihood). Without smoothing
$(\lambda=0)$, which is not shown in the plot, $\widehat{h}^2 = 0.23$.
Smoothing the relationship matrix results in an increasing estimate of
the heritability which stabilizes at about 70\%. Further smoothing
beyond the range displayed leads to a numerically unstable optimization
problem and diminished likelihood. Using the profile likelihood
approach, $\lambda$ is chosen to be the point at which REML is
maximized. This method results in an estimate of $\widehat{\lambda
}=0.20$ corresponding to $\widehat{h}^2=0.71$. Smoothing using our SNP
subsampling scheme results in $\widehat{\lambda}=0.18$ and $\widehat{h}^2=0.72$.

For comparison, we have repeated the above experiments with an
orthogonal basis computed by principal component analysis (PCA) in lieu
of a treelet basis. Such an approach does not improve the estimates of
family relationships or heritability. When noise is present, PCA is
unable to uncover the underlying sparse structure of the relationship
matrix. In fact, the results with PCA are identical to those without
smoothing (with the profile likelihood peaking when the tuning
parameter is set to 0).

Another trait that was measured in this study is the Abdomen Visceral
Fat Density (AVFD). As was the case with BMI, we restricted our
attention to individuals of European descent and regressed out age, sex
and collection site. According to the literature, the heritability of
AVFD should be between 45--70\% [\citet{Katzmarzyk1999avfd}]. According
to Figure~\ref{figBMIAVFD}, one can see that using the smoothing
parameter based on our subsampling scheme ($\widehat\lambda= 0.18$) we
get $\widehat{h}^2 = 0.29$. On the other hand, exploiting the profile
likelihood plot results in $\widehat\lambda= 0.09$ and $\widehat{h}^2 =
0.36$. When no smoothing was used (not shown in figure), $\widehat
\lambda= 0.11$. Thus, both methods for choosing the smoothing parameter
used in TCS resulted in estimates of the heritability that are closer
to what is established in the literature than without smoothing.

It is notable that $\widehat{h}^2$ for both traits increased toward
the established estimate of heritability regardless of how we estimate
the optimal smoothing parameter, because only a small fraction of the
genome was sampled by the SNP panel. Thus, our results underscore the
fact that the quantitative trait model given in equation~(\ref
{eqqtmodel}) does not require measurement of the causal SNPs that
constitute equation~(\ref{eqSNPs}). What is required is a good estimate
of $A$ based on relatives.

Our analysis of BMI and AVFD illustrates the difference between
estimates of heritability in the traditional sense and estimates of
$h^2_s$, the heritability attributable to the SNPs in the panel. From
equations~(\ref{eqknownZc}) and (\ref{equnknownZc}) it is clear that
heritability derived from the classic quantitative traits model can
distinguish between variance explained by relatives and variance
explained by causal SNPs only if either (i) all causal SNPs are
excluded, or (ii) all relatives are excluded. Because a large number of
undiscovered SNPs scattered across the genome are likely to be causal,
and large samples invariably contain distantly related individuals,
some ambiguity will always be present.

Clearly, the $90\mbox{,}000$ SNPs in our panel do not explain a substantial
fraction of the variation in BMI and yet we obtain an accurate estimate
of heritability using TCS. The increase in estimated heritability of
BMI from 23\% to 72\% suggests that smoothing improves the estimate of
$A$ and that a substantial fraction of the correlation in BMI in our
sample is due to genetic relatedness. In a similar study with a larger
population sample \citet{yang2011genome} estimated $h^2_S$ of BMI at
17\% when using the full SNP panel, but excluding all detectable
relatives. Assuming relatives were successfully removed, they conclude
that approximately 17\% of the variability in BMI is explained by
common variants included or tagged by the SNP panel.

\section*{Discussion}

Recently, there has been an upsurge of papers on sparse covariance
matrix estimation; see \citet{bickel2008regularized}, \citet
{cai2011adaptive} and the references within. Most of this research
concerns the problem of estimating population covariance matrices from
samples of multivariate data in the ``large $p$--small~$n$'' regime using
banding or thresholding techniques in the original coordinate system.
Our setting is slightly different with a more complex data structure:
We want to improve estimates of a large covariance matrix ($A$) in
which we expect a hierarchical block structure due to clustering of
distantly related individuals. A noisy estimate of covariance is
obtained from a large sample of SNPs, each of which contains very
little information. This matrix is interpreted as the additive genetic
relationship matrix and it can be used to infer degree of relationship
between pairs of individuals.

We propose a new method, which we call treelet covariance smoothing
(TCS), for regularizing real symmetric matrices with hierarchical block
structure and unstructured noise. We show how a subsampling strategy
applied to SNPs can be used to choose the tuning parameter for the
smoothing procedure. For simulated data, we show that TCS does indeed
improve estimates of family relationships. As an application we show
how TCS can be used to estimate heritability of quantitative traits
from a genome-wide sample of SNPs by smoothing relationships estimated
from those SNPs. We then apply TCS to the problem of estimating the
heritability of body mass index (BMI) and abdomen visceral fat density
(AVFD) in the Health ABC data set. In particular, BMI heritability is
usually quoted to be at least 0.50, but an estimate based on a noisy
estimate of $A$ yields a much lower value of $0.23$. By denoising the
estimated relationship matrix with treelets, we increase the estimated
heritability of BMI from $0.23$ to $0.72$. AVFD heritability analysis
produces similar results. Thus, a careful examination of heritability
estimates using more distant relatives demonstrates that one may
substantially improve relationship estimates using TCS.

Other covariance regularization schemes exist in the literature, but
systematic comparison is beyond the scope of this work. Direct
application of regularization methods for a sample covariance matrix
($Z_cZ_c^t$) is sometimes further complicated if we do not have direct
access to the multivariate data matrix $Z_c$. \citet{cai2011adaptive},
for example, describe a state-of-the-art adaptive thresholding method
for heteroscedastic problems that requires an estimate of the
variability of the entries of a sample covariance matrix. To our
knowledge, TCS is the only principled approach to regularization of
general similarity matrices with block structure on multiple scales. In
addition, the computed basis vectors themselves contain information of
the internal structure of the data---a topic that we will explore in a
separate paper with applications to complex extended pedigrees. One can
also easily modify the TCS algorithm so that positive semi-definiteness
is always guaranteed.

Our results are relevant to a recent area of burgeoning interest in
genetics, namely, the estimation of heritability from population
samples [\citet{yang10}]. However, our purpose is to estimate
heritability, as traditionally defined, rather than to determine the
fraction of variation explained by measured SNPs. We expect that the
TCS-refined genetic relationships will find wide application to other
problems in genetics, such as population-based linkage analysis [\citet
{daylinkage}], along with linear mixed models for testing association
[\citet{kang2010variance}].

Furthermore, TCS can be applied to a whole family of mixed effects
``error-in-variables'' models of the form
%
\begin{equation}
\mathbf{y} = W \bolds{\beta} + Z \mathbf{u} + \mathbf {e}, \label{eqlinearmodel}
\end{equation}
where $\mathbf{y} \in\mathbb{R}^n$ is a vector of response variables;
$\bolds{\beta} \in\mathbb{R}^p $ is a vector of fixed effects;
$\mathbf{u} \in\mathbb{R}^q $ represents random effects; and $\mathbf
{e} \in\mathbb{R}^n $ is a vector of residual errors. In the general
case, we assume that there are $c$ random effects, where each random
effect originates from a specific distribution with zero mean and
unknown variance. In vector-matrix notation,
\[
\mathbf{u}=\pmatrix{\mathbf{u}_1
\vspace*{2pt}\cr
\vdots
\vspace*{2pt}\cr
\mathbf{u}_c}
\quad  \mbox{and} \quad  Z =(Z_1,
\ldots, Z_c),
\]
where $\mathbf{u}_i$ is a $q_i \times1$ vector whose elements are the
levels of the $i$th random factor, $q=q_1 + \cdots+ q_c$, and $Z_i$ is
an $n \times q_i$ matrix of regressors for the $i$th random factor.
Assuming $\mathbb{E}(\mathbf{u})=\mathbb{E}(\mathbf{e})=\mathbf{0}$ and
\[
\operatorname{Var}\left[ %
\matrix{\mathbf{u}
\vspace*{2pt}\cr
\mathbf{e}}
\right] = \left[ \matrix{ D & 0
\vspace*{2pt}\cr
0 & E }
\right],
\]
where $D=\operatorname{diag}(\sigma_1^2 I_{q_1}, \ldots, \sigma_c^2
I_{q_c})$, yields $\mathbb{E}[\mathbf{y}]=W \beta$ and
\[
\operatorname{Var}[\mathbf{y}]= ZDZ^t + E = \sum
_{i=1}^c \sigma_i^2
Z_i Z_i^t + E,
\]
where the variance components $\sigma_1^2, \ldots, \sigma_c^2$ and $E$
are unknown and to be estimated. Now consider an \textit{error-in-variables} scenario in which the matrix $W$ of regressors of
fixed effects is known, but we only have \textit{noisy} estimates of some
or all of the positive semi-definite (p.s.d.) matrices $Z_i Z_i^t$
associated with the random effects. If these matrices have block
structure and the noise is unstructured, then one could potentially
improve estimates of variance components by first applying TCS. In our
application, for example, we looked at a special case where we first
estimate the p.s.d. matrix $Z_cZ_c^t$ in an additive polygenic model using
marker-based data, and then use a denoised estimate of $Z_cZ_c^t$ to
estimate the variance components, $\sigma_g^2$ and $\sigma_e^2$ in a
random effects model where $D=\sigma_g^2I$ and $E=\sigma_e^2 I$.

In summary, we have introduced a new method, called Treelet Covariance
Smoothing (TCS), that regularizes a relationship matrix estimated from
a large panel of genetic markers. In the context of a GWAS study a huge
number of SNPs are measured, each of which provides information about
the relationship between individuals in the sample. We proposed a SNP
subsampling procedure that exploits this rich source of information to
choose a tuning parameter for the algorithm. We illustrated one
instance of the utility of such estimates by substituting the resulting
smoothed relationship matrix into a random effects model to estimate
the heritability of body mass index. While others have used genetically
inferred estimates of relatedness from samples of close relatives to
estimate heritability, we believe this is the first time such estimates
have been applied to population-based samples when the goal is to
estimate heritability in the traditional sense.

\section*{Acknowledgments}
We would like to thank Daniel Weeks,
Nadine Melhem and Cosma Shalizi for comments on the manuscript,
Elizabeth Thompson for guidance in designing the simulations, and Evan
Klei for assistance with the simulations.

%



\printaddresses

\end{document}